\documentclass[aps,prl,twocolumn,showpacs,superscriptaddress]{revtex4-1}
\usepackage{graphics}
\usepackage{epstopdf}
\usepackage{bm}

\begin{document}
\newcommand\Hbar{$\bar{\rm H}$}
\newcommand\Hbars{$\bar{\rm H}\,$s}
\newcommand\pbar{$\bar{\rm p}$}
\newcommand\pbars{$\bar{\rm p}\,$'s}
\newcommand\pos{e$^+$}
\newcommand\poss{e$^+\,$'s}
\newcommand\elec{e$^-$}
\newcommand\elecs{e$^-$'s}
\newcommand\us{$\,$}
\newcommand\rhat{{\hat \mathbf{r}}}
\newcommand\thetahat{{\hat{\bm\theta}}}
\newcommand\zhat{{\hat\mathbf{z}}}
\newcommand\xhat{{\hat\mathbf{x}}}
\newcommand\yhat{{\hat\mathbf{y}}}

\newcommand\Lp{L_{\bar{\rm p}}}
\newcommand\npbar{n_{\bar{\rm p}}}
\newcommand\nelec{n_{{\rm e}^{-}}}
\newcommand\mpbar{m_{\bar{\rm p}}}

\hyphenation{ALPHA}

\title{Centrifugal separation and equilibration dynamics in an electron-antiproton plasma}
\author{G.B. Andresen}
\affiliation{Department of Physics and Astronomy, Aarhus University, DK-8000 Aarhus C, Denmark}
\author{M.D. Ashkezari}
\affiliation{Department of Physics, Simon Fraser University, Burnaby BC, Canada V5A 1S6}
\author{M. Baquero-Ruiz}
\affiliation{Department of Physics, University of California at Berkeley, Berkeley, CA 94720-7300, USA}
\author{W. Bertsche}
\affiliation{Department of Physics, Swansea University, Swansea SA2 8PP, United Kingdom}
\author{P.D. Bowe}
\affiliation{Department of Physics and Astronomy, Aarhus University, DK-8000 Aarhus C, Denmark}
\author{E. Butler}
\affiliation{Department of Physics, Swansea University, Swansea SA2 8PP, United Kingdom}
\author{C.L. Cesar}
\affiliation{Instituto de F\'{i}sica, Universidade Federal do Rio de Janeiro, Rio de Janeiro 21941-972, Brazil}
\author{S. Chapman}
\affiliation{Department of Physics, University of California at Berkeley, Berkeley, CA 94720-7300, USA}
\author{M. Charlton}
\affiliation{Department of Physics, Swansea University, Swansea SA2 8PP, United Kingdom}
\author{A. Deller}
\affiliation{Department of Physics, Swansea University, Swansea SA2 8PP, United Kingdom}
\author{S. Eriksson}
\affiliation{Department of Physics, Swansea University, Swansea SA2 8PP, United Kingdom}
\author{J. Fajans}
\affiliation{Department of Physics, University of California at Berkeley, Berkeley, CA 94720-7300, USA}
\author{T. Friesen}
\affiliation{Department of Physics and Astronomy, University of Calgary, Calgary AB, Canada T2N 1N4}
\author{M.C. Fujiwara}
\affiliation{TRIUMF, 4004 Wesbrook Mall, Vancouver BC, Canada V6T 2A3}
\author{D.R. Gill}
\affiliation{TRIUMF, 4004 Wesbrook Mall, Vancouver BC, Canada V6T 2A3}
\author{A. Gutierrez}
\affiliation{Department of Physics and Astronomy, University of British Columbia, Vancouver BC, Canada V6T 1Z4}
\author{J.S. Hangst}
\affiliation{Department of Physics and Astronomy, Aarhus University, DK-8000 Aarhus C, Denmark}
\author{W.N. Hardy}
\affiliation{Department of Physics and Astronomy, University of British Columbia, Vancouver BC, Canada V6T 1Z4}
\author{M.E. Hayden}
\affiliation{Department of Physics, Simon Fraser University, Burnaby BC, Canada V5A 1S6}
\author{A.J. Humphries}
\affiliation{Department of Physics, Swansea University, Swansea SA2 8PP, United Kingdom}
\author{R. Hydomako}
\affiliation{Department of Physics and Astronomy, University of Calgary, Calgary AB, Canada T2N 1N4}
\author{S. Jonsell}
\affiliation{ Fysikum, Stockholm University, SE-10691, Stockholm, Sweden}
\author{N. Madsen}
\affiliation{Department of Physics, Swansea University, Swansea SA2 8PP, United Kingdom}
\author{S. Menary}
\affiliation{Department of Physics and Astronomy, York University, Toronto, ON, M3J 1P3, Canada}
\author{P. Nolan}
\affiliation{Department of Physics, University of Liverpool, Liverpool L69 7ZE, United Kingdom}
\author{A. Olin}
\affiliation{TRIUMF, 4004 Wesbrook Mall, Vancouver BC, Canada V6T 2A3}
\author{A. Povilus}
\affiliation{Department of Physics, University of California at Berkeley, Berkeley, CA 94720-7300, USA}
\author{P. Pusa}
\affiliation{Department of Physics, University of Liverpool, Liverpool L69 7ZE, United Kingdom}
\author{F. Robicheaux}
\affiliation{Department of Physics, Auburn University, Auburn, AL 36849-5311, USA}
\author{E. Sarid}
\affiliation{Department of Physics, NRCN-Nuclear Research Center Negev, Beer Sheva, IL-84190, Israel}
\author{D.M. Silveira}
\affiliation{Atomic Physics Laboratory, RIKEN, Saitama 351-0198, Japan}
\author{C. So}
\affiliation{Department of Physics, University of California at Berkeley, Berkeley, CA 94720-7300, USA}
\author{J.W. Storey}
\affiliation{TRIUMF, 4004 Wesbrook Mall, Vancouver BC, Canada V6T 2A3}
\author{R.I. Thompson}
\affiliation{Department of Physics and Astronomy, University of Calgary, Calgary AB, Canada T2N 1N4}
\author{D.P. van der Werf}
\affiliation{Department of Physics, Swansea University, Swansea SA2 8PP, United Kingdom}
\author{J.S. Wurtele}
\affiliation{Department of Physics, University of California at Berkeley, Berkeley, CA 94720-7300, USA}
\affiliation{Lawrence Berkeley National Laboratory, Berkeley, CA 94720, USA}
\author{Y. Yamazaki}
\affiliation{Atomic Physics Laboratory, RIKEN, Saitama 351-0198, Japan}
\affiliation{Graduate School of Arts and Sciences, University of Tokyo, Tokyo 153-8902, Japan}
\collaboration{ALPHA Collaboration}
\noaffiliation

\date{Received \today}

\begin{abstract}
Charges in cold, multiple-species, non-neutral plasmas separate radially by mass, forming centrifugally-separated states.  Here, we report the first detailed measurements of such states in an electron-antiproton plasma, and the first observations of the separation dynamics in any centrifugally-separated system.  While the observed equilibrium states are expected and in agreement with theory, the equilibration time is approximately constant over a wide range of parameters, a surprising and as yet unexplained result.  Electron-antiproton plasmas play a crucial role in antihydrogen trapping experiments.

\end{abstract}


\maketitle
Non-neutral (charged) plasmas held in a Penning-Malmberg trap rotate around the trap's magnetic axis. Such traps use a solenoidal field ${\bf B}$ to provide radial plasma confinement, and electrostatic fields to provide axial confinement.  The rotation results from the ${\bf E}\times {\bf B}$ drift induced by the plasma's self electric field.  If the plasma contains multiple species, the heavier will be pushed outwards by centrifugal forces.  This effect, first predicted by O'Neil \cite{onei:81}, can lead to almost complete separation of the species in sufficiently cold plasmas.  Centrifugally separated states have been observed in several laser cooled Be$^+$--ion \cite{lars:86,imaj:97,grub:01} and Be$^+$--positron  \cite{jele:03} systems.

\begin{figure}[t]
\centerline{\resizebox{6.9723cm}{!}{\includegraphics{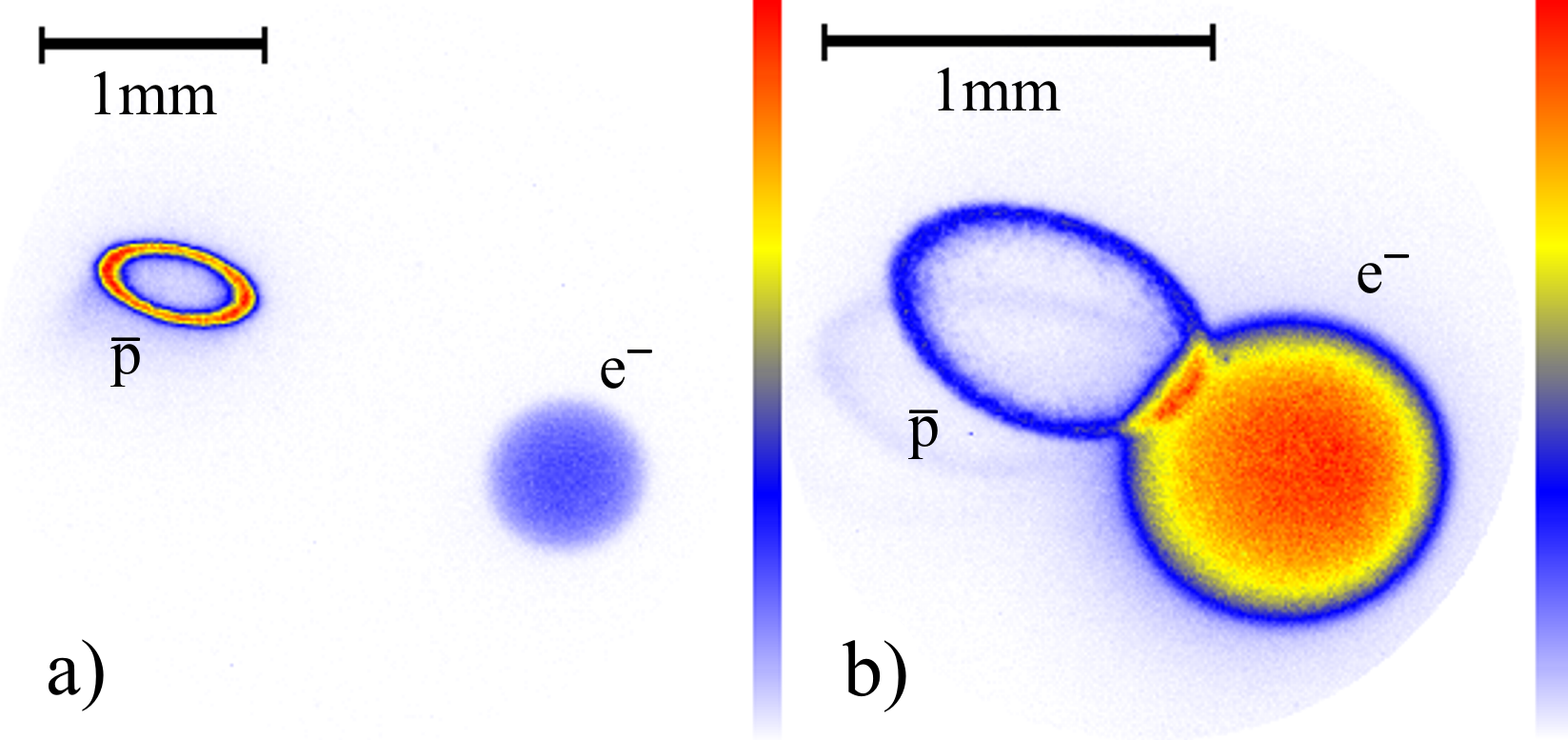}}}
\caption{(Color online) Images of centrifugally separated plasmas trapped in a) a 1\us T and b) a 3\us T solenoidal field.  In both cases the \pbar\ plasma is the ring on the left, and the \elec\ plasma is the disk on the right.    While the \pbar\ and \elec\ plasmas are concentric in the trap, they do not image to the same center.  The faint halos visible near the \pbar\ plasmas are thought to be due to optical reflections. In these and subsequent figures, the color bar is linear in density (red highest); each image is individually adjusted to saturate the color bar.  The 1\us mm length bars give the true dimensions in the trap, not the dimensions on the MCP.}
\label{Typical}
\end{figure}

This paper presents the first images and detailed measurements of centrifugal separation in an electron (\elec)--antiproton (\pbar) plasma system. Figure~\ref{Typical} shows images of two such centrifugally-separated plasmas in the ALPHA antihydrogen trapping apparatus \cite{andr:07}. (Recently, the ATRAP collaboration reported \cite{gabr:10} indirect observations of centrifugally-separated states in a \elec--\pbar\ system.)  We also report the first measurements of the separation dynamics in any centrifugally-separated system. Specifically, we report the timescale on which the \pbar\ distribution comes into equilibrium in response to changes in the \elec\ temperature and density.  Depending on the type of change in the \elec\ parameters, the \pbar\ equilibration time scale varies from milliseconds to seconds, but it is notable that the time scale is approximately seventy milliseconds for a wide range of changes. Not all of these time scales can be explained by previously explored theoretical mechanisms.


The \pbars\ will separate from the \elecs\ when the ${\bf E}\times {\bf B}$ rotational velocity $r\omega_{\rm R}$  at the plasma edge starts to exceed the thermal velocity of the \pbars. In the limit, common in this paper, that there are few \pbars\ compared to \elecs,  the \pbar\ radial density will be given by $\npbar(r,z)\sim\nelec(r,z)\exp[(r/\Lp)^2]$, where $\nelec(r,z)$ is the \elec\ radial density, and the \pbar\ edge scale length $\Lp$ is
\begin{equation}
\Lp
=\left[\frac {2 k_{\rm b} T}{\mpbar \omega_{\rm R}^2}\right]^{1/2}
=\left[\frac{8 k_{\rm b} T \epsilon_0^2 B^2}{\mpbar \nelec^2 e^2}\right]^{1/2},
\label{Rule}
\end{equation}
where $T$ is the plasma temperature, $\mpbar$ is the \pbar\ mass, $\nelec=\nelec(0,0)$ is the central \elec\ density, and $e$ is the unit charge. The rightmost expression holds only in the long plasma limit.

Images like those shown in Fig.~\ref{Typical} are obtained destructively \cite{andr:09}.  One of the electrostatic well walls that confine the mixed plasmas is lowered, thereby allowing the plasmas to flow out of the uniform solenoidal region (1\us T or 3\us T) and onto a micro-channel plate (MCP)/phosphor screen/CCD imaging system, yielding an axially-integrated image of the plasma densities.  The MCP/phosphor is located far into the solenoid's fringing field, in a field of only 0.024\us T. Electrons follow the field lines well, but, because of their heavier mass, \pbars\ are not as tightly bound to the field lines.  We observe that initially overlapping \elecs\ and \pbars\ image to different locations; the locations depend on the originating magnetic field magnitude and the distance the particles travel to the MCP/Phosphor.  When these two factors are held fixed, the centers of the \elec\ and \pbar\ images are constant for all variations of particle number and radial profile, and are independent of the presence or absence of the other species. Further, the magnetic system is astigmatic for \pbars, and the resulting images are always elliptical. The response of the imaging system to \pbars\ is more than $100\times$ greater than it is for \elecs\ because the \pbar\ annihilations greatly enhance the response \cite{andr:09}, so the relative brightness of the \elec\ and \pbar\ images can be deceiving.  For instance, in Fig.~\ref{Typical}b there are $\sim 540\times$ as many \elecs\ than \pbars.

\begin{figure}[t]
\centerline{\resizebox{7.874cm}{!}{\includegraphics{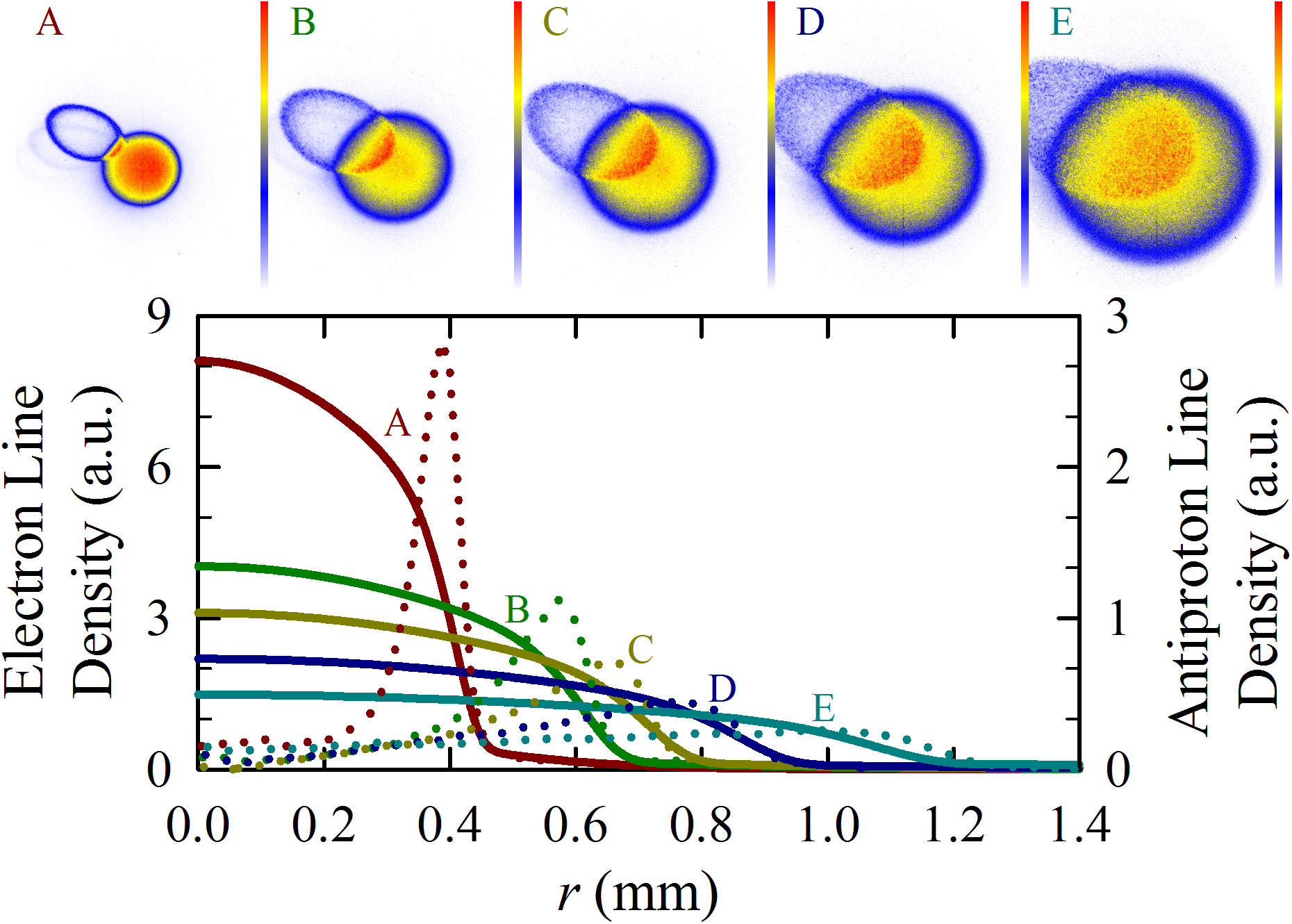}}}
\caption{(Color online) Centrifugal separation as a function of plasma size.  In all cases the solid line is the \elec\ radial profile and the dotted line is the \pbar\ radial profile. Apertures clip images D and E.
}
\label{RadialScan}
\end{figure}

\begin{figure}[t]
\centerline{\resizebox{7.8782cm}{!}{\includegraphics{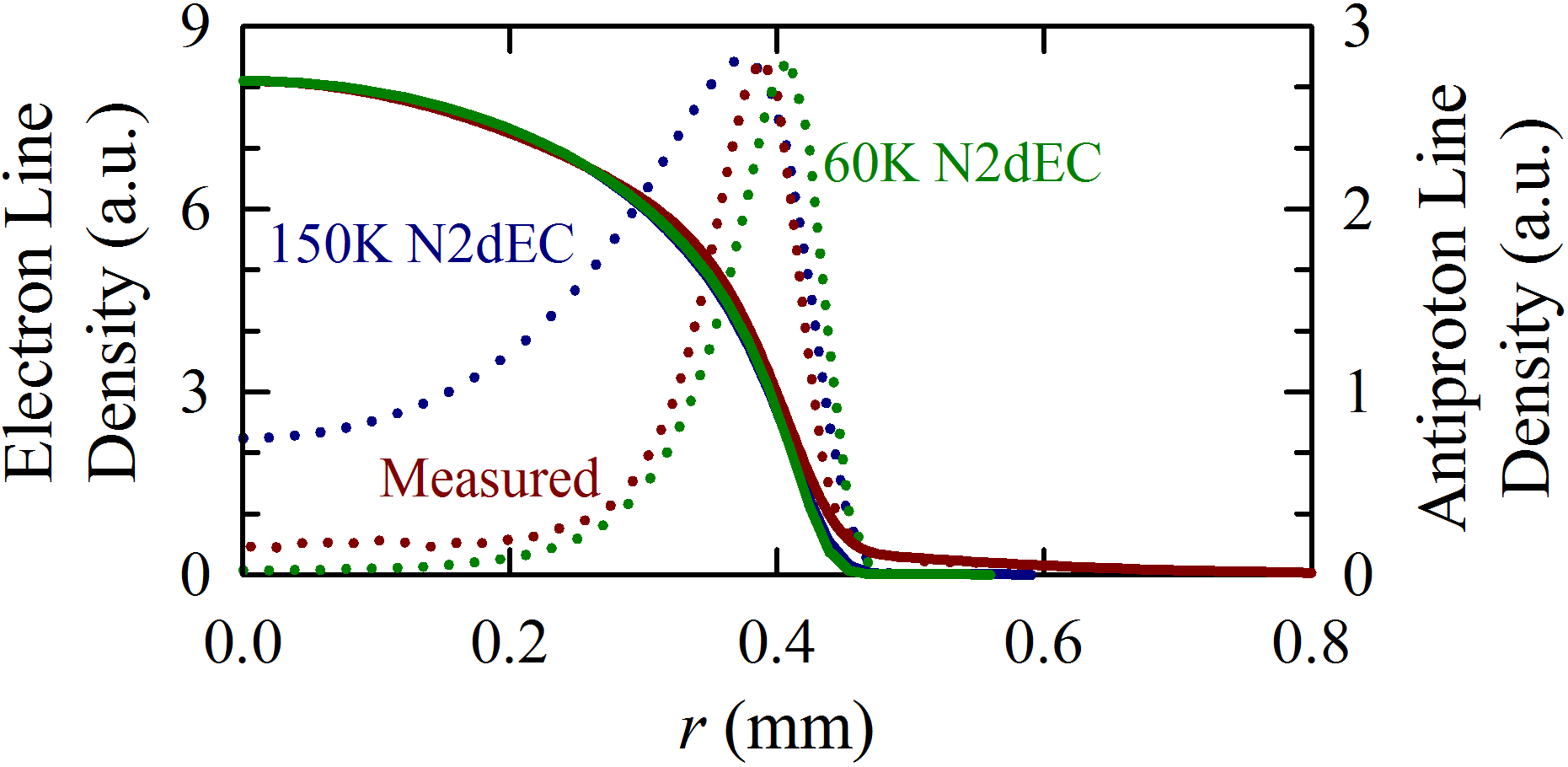}}}
\caption{(Color online) Comparison of the measured radial profiles to computed radial profiles at bracketing temperatures. The experimental conditions  are identical to profile Fig.~\ref{RadialScan}A. The solid lines are the nearly indistinguishable \elec\ radial profiles, and the dotted lines are the \pbar\ radial profiles.}
\label{RadialScanMatch}
\end{figure}

Figure~\ref{RadialScan} shows a scan of the equilibrium as a function of the radius $r_p$ of the \elec\ plasma.  In all cases, the plasmas contain approximately $1.9\times 10^7$ \elecs\ and $35,000$  \pbars, and the average temperature for the series is $120\pm 30\,\mbox{K}$ \cite{eggl:92,andr:10}. The radius of the plasma was controlled by varying the amount of radial compression \cite{andr:08} applied to the plasma before the equilibrium was established. The central \elec\ densities range from $1.9\times 10^9\,\mbox{cm}^{-3}$ to $2.7\times 10^8\,\mbox{cm}^{-3}$.  Under the constant total charge $Q$ conditions of Fig.~\ref{RadialScan}, the relative scale ratio $\Lp/r_p$  is proportional to $Br_p\sqrt{T}/Q$.  Thus, we would expect that the \elecs\ and \pbars\ would separate less as the plasma radius increases, a trend clearly observed in Fig.~\ref{RadialScan}. Our plasmas are close to ellipsoids with length to diameter ratios of 30; to more accurately estimate the degree of separation, we developed a numeric, two-dimensional ($r$--$z$) equilibrium code (N2dEC) which includes the necessary finite-length effects \cite{lars:86,imaj:97,grub:01}.  Figure~\ref{RadialScanMatch} replots the smallest radius plasma from Fig.~\ref{RadialScan}, and the results of N2dEC calculations at bracketing temperatures.  As the measured temperatures are subject to systematic and shot-by-shot errors, it is not surprising that the best-fit temperature demanded by the code does not perfectly match the measured temperature.

The line densities and (in later figures) moments used throughout this paper are computed from one-dimensional fits to the two dimensional images from the MCP/Phosphor.
The fits also correct for the ellipticity of the \pbar\ images; charge along all points on any given ellipse is mapped to the major radius of that ellipse.

\begin{figure}[t]
\centerline{\resizebox{7.6539cm}{!}{\includegraphics{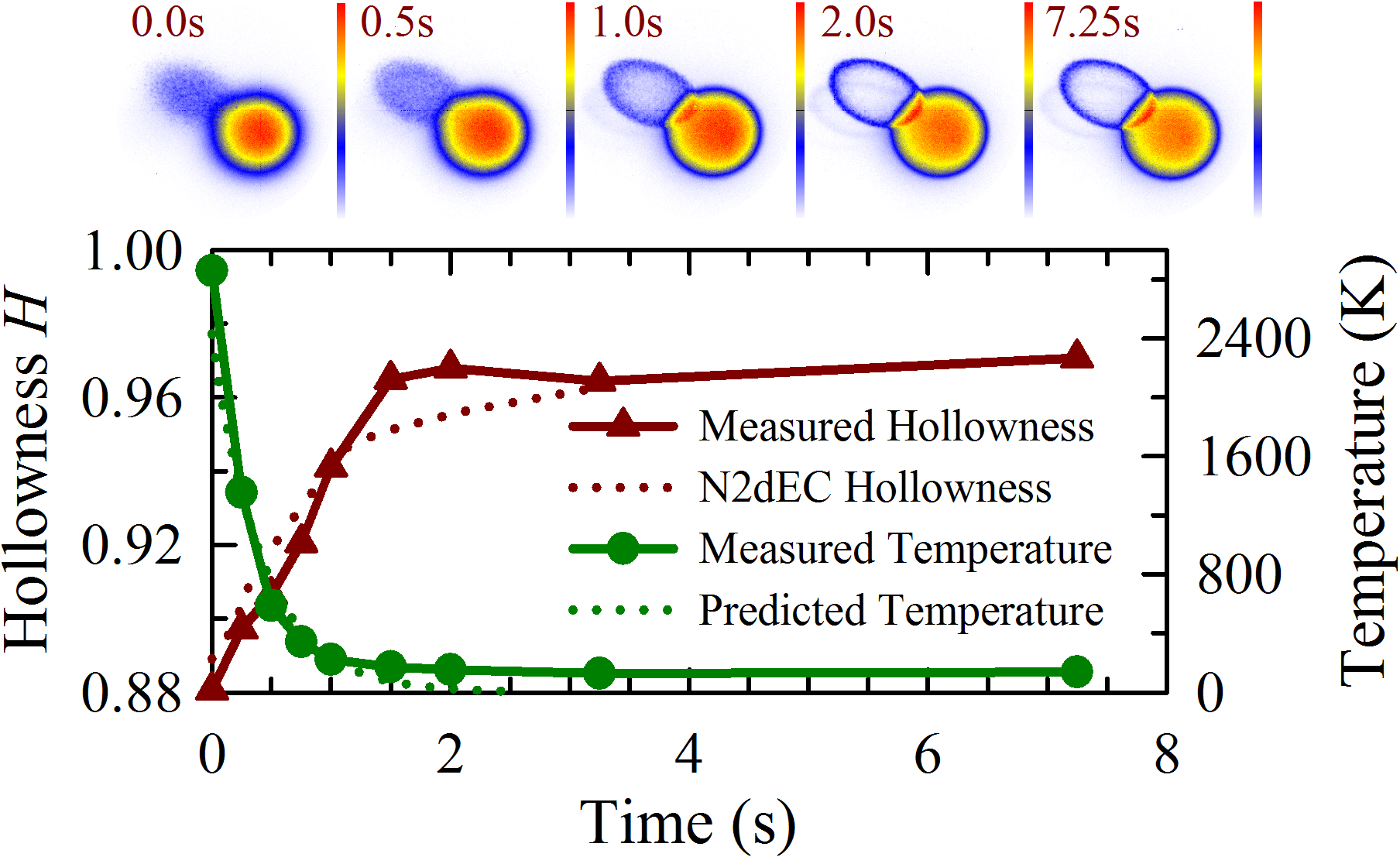}}}
\caption{(Color online) Temperature and hollowness $H$ as a function of time in a cooling plasma for a plasma similar to that in Fig.~\ref{Typical}b. The predicted temperature is fit through the data at 0.25\us s.}
\label{TemperatureScan}
\end{figure}

The \elecs\ in our plasmas cool via cyclotron radiation with a calculated e-folding time of 0.427\us s at 3\us T \cite{beck:92}; the \elecs\ cool the \pbars\ by collisions.  If we actively heat a mixed \elec--\pbar\ plasma, we can watch the centrifugal separation reemerge as the plasma cools.  Figure~\ref{TemperatureScan} shows the measured and predicted temperature as a function of time in such an experiment.  The measured temperatures follow the predicted temperatures until they level off, for unknown reasons, at about 130\us K. Figure~\ref{TemperatureScan} also shows the measured hollowness $H$ of the \pbar\ radial profile as a function of time, and $H$ predicted by N2dEC calculations at the measured temperatures.  This metric is defined by
$H= {\langle r^2\rangle^{0.5}}  /  {\langle r^4\rangle^{0.25}}$,
where $\langle\,\rangle$ denotes an average over the radial profile.  $H=0.841$ for a Gaussian profile, $0.915$ for a constant density ellipsoid, and $1$ for a thin annulus. The value of $H$ is robust to many forms of experimental noise, and does not depend on the overall plasma radius, but is subject to a small offset if the background in the images is imperfectly subtracted.
Figure~\ref{TemperatureScan} shows that the \pbar\ hollowness $H$ follows the plasma temperature, suggesting that the \pbars\ go through a series of quasi-equilibrium states. No delay in attaining equilibrium is visible.

\begin{figure}[t]
\centerline{\resizebox{7.2729cm}{!}{\includegraphics{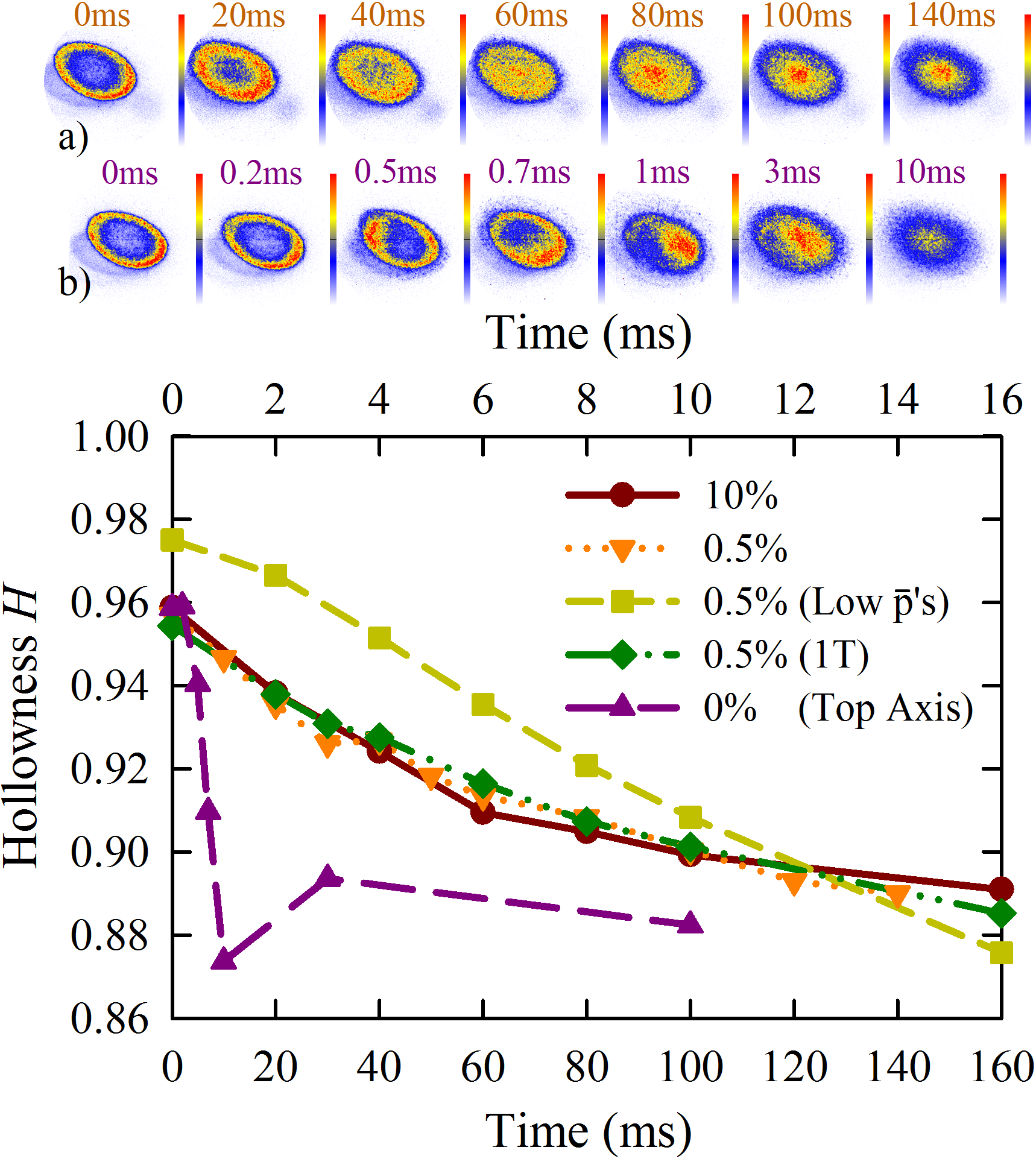}}}
\caption{(Color online) Hollowness as a function of time after e-kicking leaves the stated percentages of electrons in the plasma. The state before e-kicking is similar to that shown in Fig.~\ref{Typical}b.  Images a) correspond to the 0.5\% series. The \elec plasma is the faint disk visible in the lower right of the images (but remember that there are still more \elecs\ than \pbars.) The apparent smallness of the \elec\ plasma is an artifact of the faintness of the image. Images b) correspond to the 0\% series.}
\label{E-KickScan}
\end{figure}

If we change the \elec\ conditions suddenly, the \pbars\ must relax to the new equilibrium corresponding to these new conditions.  Figure~\ref{E-KickScan} shows a study of this process when a specified fraction of the electrons is suddenly removed.  This is accomplished by lowering an electrostatic well sidewall for a time short ($<1\,\mu\mbox{s}$) compared to the time over which the \pbars\ can move, but long enough for the required fraction of \elecs\ to escape.  This process, called ``e-kicking'', substantially heats the remaining \elecs\ to several thousand Kelvin.  Several cases were studied: leaving 10\% of the \elecs, so that the density of \elecs\ is $3.3\times 10^8\,\mbox{cm}^{-3}$; leaving 0.5\% of the \elecs, so that the density is $5.3\times 10^7\,\mbox{cm}^{-3}$; leaving 0.5\% of the \elecs, but with $3,100$ \pbars\ rather than $35,000$ \pbars\ as in all the other data; leaving 0.5\% of the \elecs, but at 1\us T rather than at 3\us T as in all the other data; and leaving 0\% of the \elecs.

The e-kicking process leaves the \pbars\ in their initial, ring-like distribution, but, because the \elec\ density is now low and the temperature now high, the final equilibrium profile is Gaussian-like.  Considering the first four, non-0\% cases only, the time to re-establish equilibrium (somewhat arbitrarily set at $H=0.91$) is approximately 60\us ms (slightly longer for the low \pbar\ case.)  The images (Fig.~\ref{E-KickScan}a) for this class of relaxation show no sign of instability; the center appears to fill in gradually and uniformly.  The last case, in which all the \elecs\ are removed, is very different.  The relaxation is approximately $10\times$ faster, and proceeds (Fig.~\ref{E-KickScan}b) via a classic $\ell=1$ diocotron instability \cite{dris:90a}. Three-dimensional molecular dynamics simulations  verify that the plasmas are indeed $\ell=1$ unstable under these conditions.  In the other cases, leaving even as few as 95,000 \elecs\ (the 0.5\% case) reduces the rotational shear to below the threshold necessary to drive the $\ell=1$ instability.

\begin{figure}[t]
\centerline{\resizebox{8.3693cm}{!}{\includegraphics{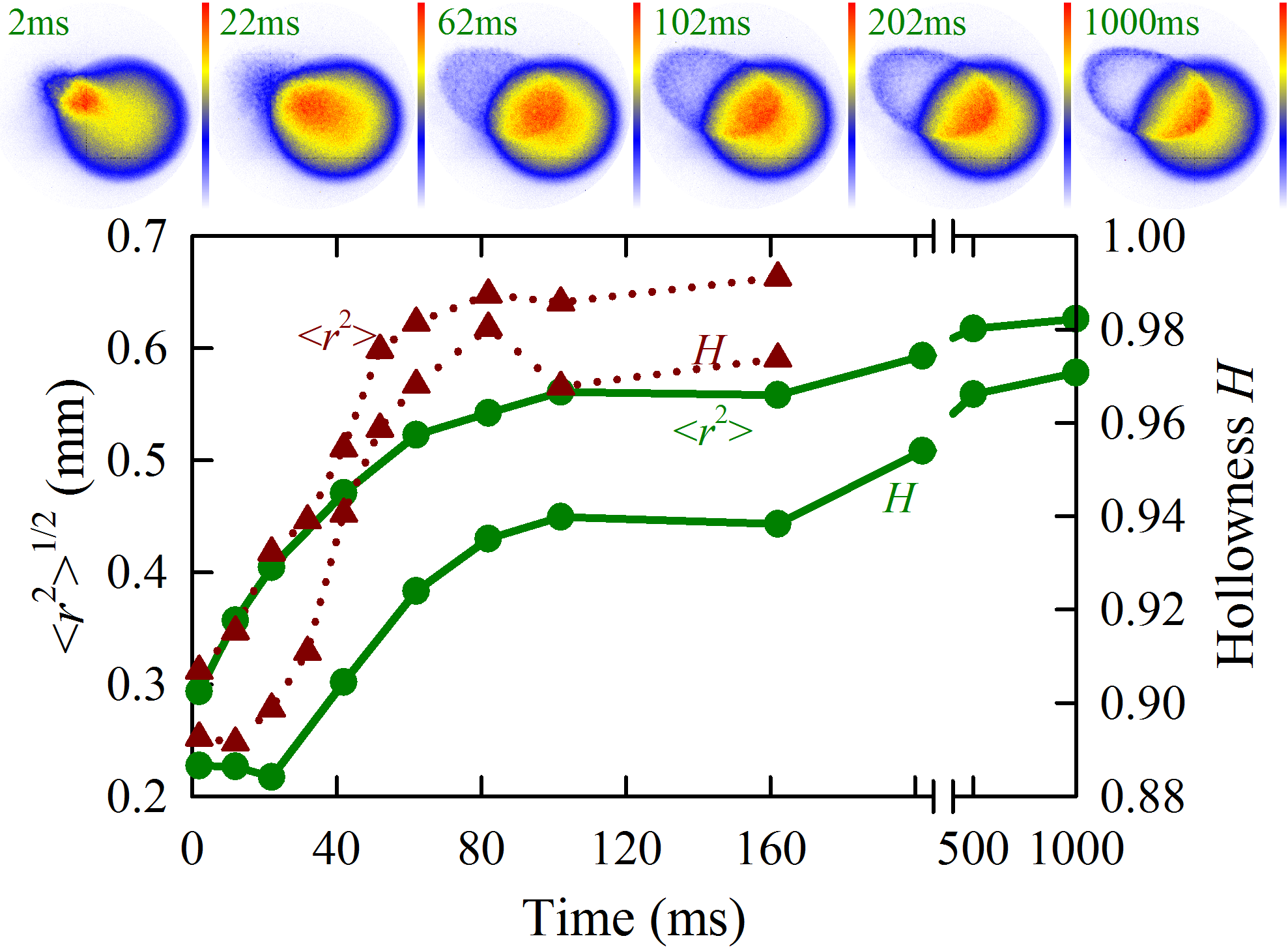}}}
\caption{(Color online) Radius and hollowness as a function of time after injection of $35,000$ \pbars\ (circles) and $8,800$ \pbars\ (triangles).  The top set of images corresponds to the $35,000$ \pbars\ series.}
\label{SlugInjection}
\end{figure}

We have also studied the injection of a very small radius \pbar\ ``slug'' into a pre-existing, larger radius, cold \elec\ plasma.  We then track two distinct measures: the expansion of the \pbar\ slug out to the radius of the \elecs, and the subsequent hollowing of the \pbars.  Figure~\ref{SlugInjection} shows two such series, corresponding to the injection of a standard load of \pbars\ ($35,000$) into a standard \elec\ plasma (Fig.~\ref{Typical}b-like, but with an \elec\ density of $1.5\times 10^9\,\mbox{cm}^{-3}$), and injection of a reduced load of \pbars\ ($8,800$) into the same \elec\ plasma.  The \pbars\ begin to expand immediately, reaching a plateau after approximately 80\us ms. Hollowing begins after about 20\us ms, and plateaus shortly after the plasma stops expanding. Injecting the \pbar\ slug heats the \elecs, particularly for the standard \pbar\ load where the increase in temperature is approximately 150\us K.  The late-time evolution ($t>200\,\mbox{ms})$ visible in the standard \pbar\ load series is probably due to the \elec\ plasma radiating away the energy brought in during the \pbar\ injection.  The reduced \pbar\ load heats the \elec\ plasma less (approximately 50\us K); consequently, it becomes hollow somewhat faster.

In conclusion, the \pbars\ in an initially hot \elec\ plasma appear to follow an evolving equilibrium as the \elec\ cool.  The evolution time scale is set by the cooling timescale.  When all the \elecs\ supporting a well-developed \pbar\ ring equilibrium are removed (Fig.~\ref{E-KickScan}b), the ring collapses via an $\ell=1$ diocotron instability on a time scale of a few milliseconds.  In all other observed cases (Fig.~\ref{E-KickScan}), including the injection of a \pbar\ slug into a cold \elec\ plasma (Fig.~\ref{SlugInjection}), equilibration takes on the order of 60--80\us ms.  These cases range in \elec\ density over a factor of nearly 30, in \elec\ temperature over more than a factor of 10, in magnetic field over a factor of 3, and in \pbar\ number over a factor of 11.  The relevant collision frequencies, gyro radii, Debye lengths, etc.\ will vary over a proportionate range.  The equilibration time scale for the higher density, lower temperature data shown in Fig.~\ref{SlugInjection} is compatible with the predictions of particle diffusion/mobility theory \cite{DubinPrivate}, but the data in Fig.~\ref{E-KickScan} are not; the densities for these cases are too low, and the temperatures too high.  Test particle calculations \cite{amor:06b} predict equilibration times that are too slow by several orders of magnitude, even for the data shown in Fig.~\ref{SlugInjection}. A more exotic theory appears to be required, perhaps like that in Ref.~\cite{dubi:10a}.

 The results presented here are relevant to several of the processes employed in antihydrogen (\Hbar) trapping experiments at CERN \cite{andr:07,gabr:08}.  For example, \elecs\ are used to sympathetically cool the \pbars, but are generally removed (e-kicked) before synthesis of \Hbar\ is initiated.  Our measurements are the first detailed exploration of the radial and temporal dynamics of the \pbar\ equilibration during this process.  We have found it necessary to minimize the radius of the \pbars\ to
 enhance the probability of trapping \cite{andr:10b}; we do this by compressing the \elecs\ in a still mixed \elec--\pbar\ plasma, and relying on equilibration between the two species to compress the \pbars\ \cite{andr:08}.  Compressing the \elecs\ too quickly leaves some \pbars\ behind; the measurements here give a lower bound for the compression scale time.  Finally, in some proposed mixing schemes \cite{hu:07}, the \pbars\ and \elecs\ are left unseparated;  the effects of centrifugal separation must be considered in these schemes.

This work was supported by CNPq, FINEP/RENAFAE (Brazil), ISF (Israel), MEXT (Japan), FNU (Denmark), VR (Sweden), NSERC, NRC/TRIUMF AIF FQRNT(Canada), DOE, NSF (USA), and EPSRC, the Royal Society and the Leverhulme Trust (UK). We thank D.H.E. Dubin for his helpful comments, and S. Kemp and C. {\O} Rasmussen.


%

\end{document}